\documentclass[aps,prl,preprint]{revtex4}

\usepackage{graphicx}
\usepackage{dcolumn}
\usepackage{bm}
\usepackage{amsmath}
\begin{document}

{\title{Relation of Curie temperature and conductivity: 
(Ga,Mn)As alloy as a case study}
\author{J. Kudrnovsk\'y}
\affiliation{Institute of Physics, Academy of Sciences of the
Czech Republic, Na Slovance 2, CZ-182 21 Praha 8, Czech Republic}

\author{G. Bouzerar}

\affiliation{Institut N\'eel, MCBT CNRS, 25 rue des Martyrs, 
F-38 042 Grenoble, France}

\author{I. Turek}
\affiliation{Institute of Physics of Materials, Academy of Sciences 
of the Czech Republic, \v{Z}i\v{z}kova 22, CZ-61662 Brno, Czech Republic} 

\date{\today}

\begin{abstract}
Experimental investigations of diluted magnetic semiconductors
indicate a strong relation between Curie temperature and conductivity.
Both quantities depend non trivially on the concentration of magnetic
impurities, the carrier density, and the presence of compensating defects.
We calculate both Curie temperature and conductivity of (Ga,Mn)As alloys
in a selfconsistent manner based on the same first principles Hamiltonian
in which the presence of compensating defects is taken into account.
The effect of As-antisites and Mn-interstitials is determined separately
and a good agreement between theory and experiment exists only in the
case where the dominating mechanism of is due to the Mn-interstitials.
\end{abstract}


\maketitle
 
The diluted magnetic semiconductors (DMS) represent a promising material
for future applications in the spintronics as well as a system on
which existing theories of finite-temperature magnetism and transport
in complex diluted magnetic alloys can be critically tested.
We refer the reader to a recent extensive review \cite{review} of 
both theoretical and experimental efforts in this field.
The best studied system up to now is (Ga,Mn)As for which reliable
structural, magnetic, and transport data are now available.
Samples with well defined structural and defect characterization are
necessary requirement for any theoretical study, in particular that 
based on parameter-free approaches where quantitative agreement is
one of the primary aims.

The Curie temperature (T$_{c}$) and the conductivity are among the most 
important characteristics of magnetic alloys and their parameter-free 
determination is thus of a great importance.
Their internal relation was the subject of a recent experimental study 
\cite{edmonds}. 
A qualitative understanding of finite-temperature magnetism including
an estimate of the Curie temperature \cite{dietl} as well as of transport 
properties \cite{jungw} can be obtained using the perturbative 
J$_{pd}$-model \cite{review}.
However, this approach fails quantitatively. 
Indeed, T$_{c}$ is evaluated in the framework of the mean-field theory 
using the averaged lattice model \cite{dietl} and thus explicitly neglecting
the disorder in the system.
The conductivity is calculated in the lowest order of the perturbation 
theory which underestimates the effects of the disorder.

In the present paper we wish to calculate T$_{c}$ and the conductivity starting 
from the same, first principles Hamiltonian.
This allows us to evaluate exchange integrals as well as transport properties 
on equal footings.
A quantitative comparison of results of the parameter-free approach with 
corresponding experimental data will represent a crucial test of existing 
theories. 
First, the Curie temperature is determined beyond the average lattice 
approximation as concerns the effects of compensation and dilution and beyond 
the mean-field approach as concerns the effect of spin fluctuations 
\cite{lars,sato,epl}.
Second, the conductivity is evaluated using the Kubo-Greenwood formula which 
represents a non-perturbative approach.
We stress that the present approach has no adjustable parameters. 
Indeed, we will use for each sample the measured nominal Mn concentration 
and the carrier density which will be then used to estimate the concentration 
of compensating defects.
We will consider, as compensating defects, separately As-antisites
and Mn-interstitials and demonstrate a key relevance of the latter
defects for a good quantitative agreement with experimental data 
\cite{edmonds}.
 
We will now briefly describe methods used to calculate the
Curie temperatures and conductivities for samples at various
states of annealing corresponding to a given nominal Mn-concentration.
 
We evaluate the Curie temperature in the spirit of a two-step model as 
suggested by Liechtenstein \cite{lie}.
In the first step are total energies of low-lying magnetic excitations 
mapped onto the classical random Heisenberg model in which are the
pair exchange integrals determined explicitly from energies corresponding 
to small rigid rotations of spins on the lattice in the framework of 
the adiabatic approximation by employing the Green function approach. 
In the second step, methods of statistical physics are used to estimate 
the Curie temperature for this Hamiltonian. 
This approach has proved to be very useful for a broad class of 
magnetic materials \cite{eirev,ourJ}, including the DMS 
\cite{lars,sato,epl}.
We have implemented above approach in the framework of the 
all-electron scalar-relativistic tight-binding linear muffin-tin 
orbital (TB-LMTO) method in which the disorder due to various (small)
defect concentrations is included in the framework of the coherent 
potential approximation (CPA) \cite{book}.
The CPA is known to reproduce reliably the compositional dependence
of carrier concentrations as well as to describe correctly the
transport relaxation time due to the presence of various defects
in random systems.
It should be noted that the same calculated carrier concentration
and carrier lifetimes are used for the evaluation of exchange
integrals and the conductivity in the present approach in a striking 
contrast to the J$_{pd}$-model where carrier lifetime is neglected
and has to be implemented additionally for the evaluation of the
conductivity.
On the other hand, the CPA cannot describe properly behavior of
electrons in localized states in which, for example, the 
hopping conductivity dominates. 
Here, however, we consider systems in the metallic regime.

The Curie temperature as determined from the Heisenberg Hamiltonian
is due to transversal spin fluctuations neglecting thus the effect
of longitudinal (Stoner) excitations. 
This is a good approximation for systems with large rigid magnetic
moments such as Mn-impurities in GaAs semiconductor \cite{eirev}. 
The main problem of reliable determination of T$_{c}$ is the
dilution, namely the fact that for low defect concentrations
typical for DMS is the occurrence of long-range ferromagnetism
strongly depending on the spatial extent of exchange integrals.
This is particularly important for the present parameter-free
approach in which exchange integrals are not fitted but their values
and spatial extent are determined by underlying electronic structure.
Recently, three groups have elaborated independently an approach in 
which random positions of impurities are incorporated in the framework 
of the Monte Carlo sampling while corresponding transversal spin 
fluctuations are included either by the Monte Carlo approach
\cite{lars,sato} or by using the selfconsistent local random-phase 
approximation (SC-LRPA) \cite{epl}.
These theories have explained the reduction of T$_{c}$ in diluted
samples as compared to predictions based on the averaged lattice
model neglecting disorder \cite{alm}.
A detailed comparison with experiments for well annealed (Ga,Mn)As 
samples with very low concentrations of native defects (Mn-interstitials 
and As-antisites) has resulted in a good quantitative agreement 
between measured and calculated Curie temperatures (for the most
detailed comparison see Ref.~\onlinecite{epl}).

The next step in the quantitative comparison between the theory and
experiment was the study, Ref.~\onlinecite{mnias}, in which it
was demonstrated that Mn-interstitials rather than As-antisites
are dominating defects reducing T$_{c}$ as compared to optimally
annealed samples.
Such a model is supported by recent first-principles theoretical 
studies \cite{mnint}.
We will continue further and evaluate, for the same samples,
corresponding conductivities and compare the resulting T$_{c}$ vs
conductivity relation with experiment \cite{edmonds}.

The conductivity in (Ga,Mn)As is due to $p$-hole carriers which 
are through the $pd$-coupling scattered on defects present in the 
system, namely on Mn-impurities, Mn-interstitials, and As-antisites.
Without external magnetic field, there are essentially three
different contributions to the resistivity in (Ga,Mn)As alloys:
(i) the scattering on phonons, acoustical eventually optical
ones if they could be excited, (ii) the magnetic scattering
due to thermodynamical fluctuations which is the largest at 
T=T$_{c}$ and which is essentially determined by the spin-spin 
correlation function \cite{trspd}, and (iii) the residual resistivity 
due to the presence of various defects.

The residual resistivity clearly dominates in the low temperature
regime and in this paper we will limit ourselves to this case for
which experimental data \cite{edmonds} are also available.
Inclusion of the other above mentioned contributions is beyond
the scope of the present paper.
The linear-response theory (Kubo-Greenwood approach) \cite{kglmto}
as formulated in the framework of the TB-LMTO-CPA approach
is used to determine the residual conductivity of the sample.
It should be noted that corresponding transport relaxation time is
estimated in the framework of the same first-principles Hamiltonian 
as used for the estimate of exchange integrals.
The theory formulated for the multi-sublattice case allows to
include Mn-interstitials and As-antisites on equal footing
with conventional Mn-impurities \cite{trit}.
In addition, we have also implemented into the formalism
disorder-induced vertex corrections although their effect on
the conductivity in the preaent case is relatively small (less 
than 10\%).
It should be noted that the present linear theory is unable
to describe the hopping conductivity between states in well
localized impurity subbands such as the Mn-impurity bands in 
Mn-doped GaN alloys. 
The conductivity is rather sensitive to various kinds of defects
and therefore the sample preparation (the presence of defects)
will play an important role for calculated values of the conductivity.

The Curie temperature of both as-grown and annealed samples
of (Ga,Mn)As for a nominal concentration of Mn-atoms 
$x_{\rm Mn}$=0.067 was evaluated in Ref.~\onlinecite{mnias}.
Here we briefly reiterate the main features of the theory.
It was demonstrated that a small amount of As-antisites, e.g.,
0.5\%, will influence calculated T$_{c}$ only negligibly
(5-10~K).
On the other hand, the dominant mechanism for compensation
are Mn-interstitials and a good agreement between theory and
experiment was obtained for both as-grown and annealed samples.
Theory is based on the following simple model: Mn-interstitials
are attracted by substitutional Mn-atoms on Ga-sublattice and 
form a singlet pair of spins with a strong antiferromagnetic 
coupling \cite{mnint}. 
They have only weak influence on remaining magnetically active 
Mn-atoms with the effective concentration 
$x_{\rm eff}$=$x_{\rm Mn} - 2 x_{\rm Mn}^{i}$, where 
$x_{\rm Mn}$ and $x_{\rm Mn}^{i}$ are, respectively, the nominal 
(total) concentration of Mn-atoms and the concentration of 
Mn-interstitials.
We will thus eliminate these pairs from effective Heisenberg
model for the determination of T$_{c}$.
Further, by assuming that each Mn-interstitial adds two electrons 
(Mn-interstitial acts as a double donor), we obtain for the 
effective carrier concentration $n_{\rm eff}$ the relation 
$n_{\rm eff}$= $x_{\rm Mn} - 3 x_{\rm Mn}^{i}$ \cite{mnias}.
 In the framework of our model, we identify the experimental
compensation ration, $\gamma_{\rm exp}$, with the effective
compensation ratio $\gamma_{\rm eff}$=$n_{\rm eff}$/$x_{\rm eff}$.
 From known $\gamma_{\rm exp}=\gamma_{\rm eff}$ and the nominal 
Mn-concentration $\rm Mn$ we can determine $x_{\rm eff}$ and 
$n_{\rm eff}$ which are needed for the estimation of corresponding 
exchange integrals \cite{ourJ}.
The first-principles theory, however, does not allow an independent
variation of the concentration of active magnetic atoms and
the carrier concentration.
To overcome this problem one can adopt the rigid band model: 
a frozen electronic structure corresponding to $x_{\rm eff}$ is
used and the Fermi energy is shifted so that required carrier 
concentration is obtained.
This is, however, a non-selfconsistent procedure with obvious
limitations.
Instead, we have used effective doping by non-magnetic impurities
which only negligibly influence states at the Fermi energy and
thus also exchange integrals and thermodynamic magnetic properties.
The doping is thus used as a purely computational device. 
In Ref.~\onlinecite{mnias} we have used As-antisites as above
described computational device.
Formally, each singlet pair consisting of Mn-substitutional 
impurity on Ga-sublattice and the nearest-neighbor Mn-interstitial 
is thus replaced effectively by one half of the 
As$_{\rm Ga}$-antisite.
In this way are Mn-interstitals effectively excluded from the
thermodynamical studies.

On the other hand, all defects contribute to the impurity scattering,
and thus to the transport calculations, i.e., also magnetically 
inactive pairs of substitutional and interstitial Mn-impurities
have to be included.
It should be noted that the concentration of Mn-interstitial atoms
can also be easily determined from known nominal Mn-concentration
x$_{\rm Mn}$ and the compensation ratio $\gamma$.
While a small concentration of As-antisites has negligible effect
on values of T$_{c}$ as already mentioned, their effect on transport
properties has to be included.
We have therefore performed three sets of transport calculations:
(i) with As-antisites as the only compensating defect, (ii) assuming
Mn-interstitials as compensating defects, and (iii) including 
additionally to Mn-interstitials also a small concentration of As-antisites.  
We have then compared theoretical results with experimental T$_{c}$ vs 
conductivity data at low temperatures assuming that small concentration 
of As-antisites will not influence T$_{c}$ as discussed above. 
In transport calculations we have assumed, in an agreement with 
theoretical calculations \cite{mnint}, antiparallel orientations of 
spin moments on substitutional and interstitial Mn-atoms in (Ga,Mn)As
alloy and included disorder-induced vertex corrections.

The results are summarized in three figures, Fig.~\ref{f1} for 
the model with As-antisites as compensating defect, 
Fig.~\ref{f2} for a model with Mn-interstitials as compensating
defects, and Fig.~\ref{f3}, where, in addition to Fig.~\ref{f2},
are also As-antisites included in transport calculations.
The experimental data are for a nominal Mn-concentration 
$x_{\rm Mn}=0.067$ and various stages of annealing, ranging from
as-grown sample (low T$_{c}$ and conductivity) to an almost
perfectly annealed sample (high T$_{c}$ and conductivity). 
The theoretical data correspond to as-grown sample with 
$\gamma_{\rm eff}$=0.54, to partly annealed sample with 
$\gamma_{\rm eff}$=0.83, and to an almost perfectly annealed 
sample with $\gamma_{\rm eff}$=0.96.
This corresponds, respectively, to $x_{\rm Mn}^{i}$=0.016, 0.009,
and 0.0035.
It should be noted that the effective carrier concentrations 
$n_{\rm eff}$ and/or corresponding compensation ratios $\gamma$
are known with error-bar of order 15-20\% \cite{edmonds,tcexp}.
Above values of parameter $\gamma$ are the same as those used
in the paper \cite{mnias}.
In the model with both Mn-interstitials and As-antisites (Fig.~\ref{f3}), 
we have chosen the concentration of As-antisites $y_{\rm As}$=0.005
which is within the upper limit of uncertainty with which is known
effective carrier concentration experimentally and which only negligibly
(few K) influences the calculated Curie temperature.
There are no experimental data in Ref.~\onlinecite{edmonds} concerning
amount of As-antisites but it is believed that some native As-defects 
are present even in well annealed samples.
The experimental data used in Figs.\ref{f1}-\ref{f3} correspond 
to the low-temperature measurements of the conductivity (T=4.2~K)
but similar, almost linear dependence of T$_{c}$ vs conductivity
curve was obtained also for the room temperature \cite{edmonds},
although here are conductivities reduced by phonons and by additional
magnetic scattering due to thermodynamical fluctuations as described 
above.

The following conclusions can be made:
(i) As-antisites as compensating defects fail to reproduce
both the experimental T$_{c}$ (see also \cite{mnias}) and 
measured conductivity which seems to be too high in this model;
(ii) the presence of Mn-interstitials significantly improves 
the quantitative agreement between theory and experiment
by correctly reproducing an almost linear dependence of the
Curie temperature on the conductivity with an acceptable 
quantitative agreement.
While calculated T$_{c}$ agree with experimental data very
well (see also Ref.~\onlinecite{mnias}), the conductivity 
still seems to be slightly overestimated;
(iii) the additional presence of a small amount of As-antisites
not only reproduces the (almost) linear dependence of the Curie 
temperature on the conductivity but it also brings the theory 
into a good quantitative agreement with available experimental 
data.

We have presented  a parameter-free theory which is able to reproduce 
simultaneously both the Curie temperature and the low-temperature
conductivity measured in as-grown and annealed samples of (Ga,Mn)As. 
In addition, we have demonstrated that the inclusion of Mn interstitials 
into the theory is of crucial importance for a good quantitative agreement 
between theory and experiment.
As-antisites are shown to play a secondary role.

Authors acknowledge Dr. K.~Edmonds for sending his unpublished data
which made possible a detailed comparison between theory and 
experiment (see Ref.~\onlinecite{tcexp}).
J.K. and I.T. acknowledge the financial support from 
AVOZ10100520, the Grant Agency of the Academy of Sciences of 
the Czech Republic (A100100616), and from COST P19 (OC150).


\newpage

\begin{figure}
\center \includegraphics[width=12cm]{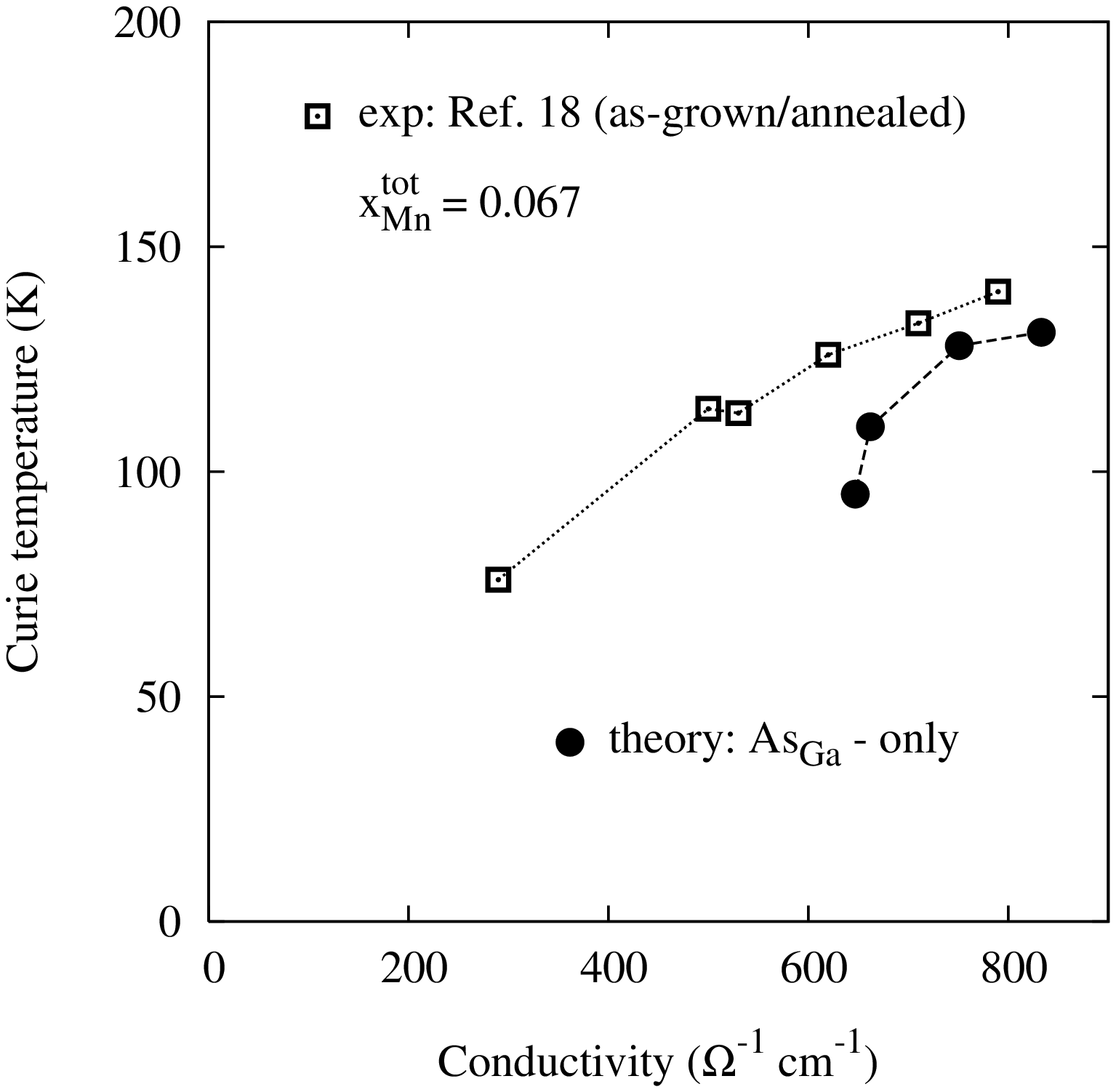}
\caption {T$_{c}$ vs conductivity relation assuming that
dominating compensating defects are As$_{\rm Ga}$-antisites.
Experimental results \cite{tcexp} for as-grown and
annealed samples are also shown.
}
\label{f1}
\end{figure}

\newpage

\begin{figure}
\center \includegraphics[width=12cm]{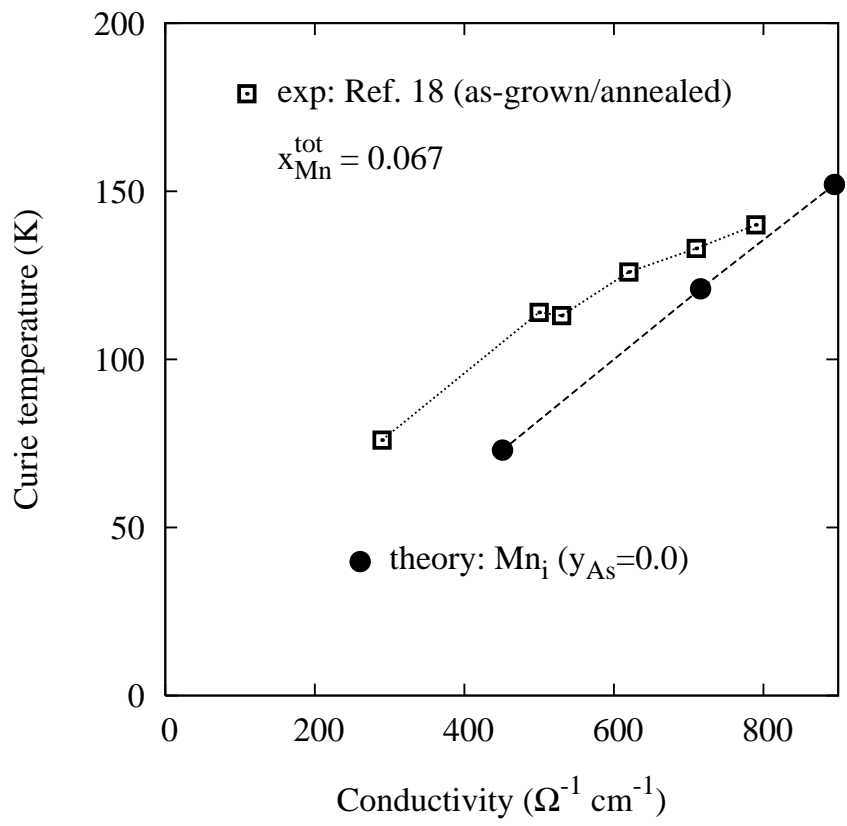}
\caption {T$_{c}$ vs conductivity relation assuming that
dominating compensating defects are Mn-interstitials. 
Experimental results \cite{tcexp} for as-grown and
annealed samples are also shown. 
}
\label{f2}
\end{figure}

\newpage

\begin{figure}
\center \includegraphics[width=12cm]{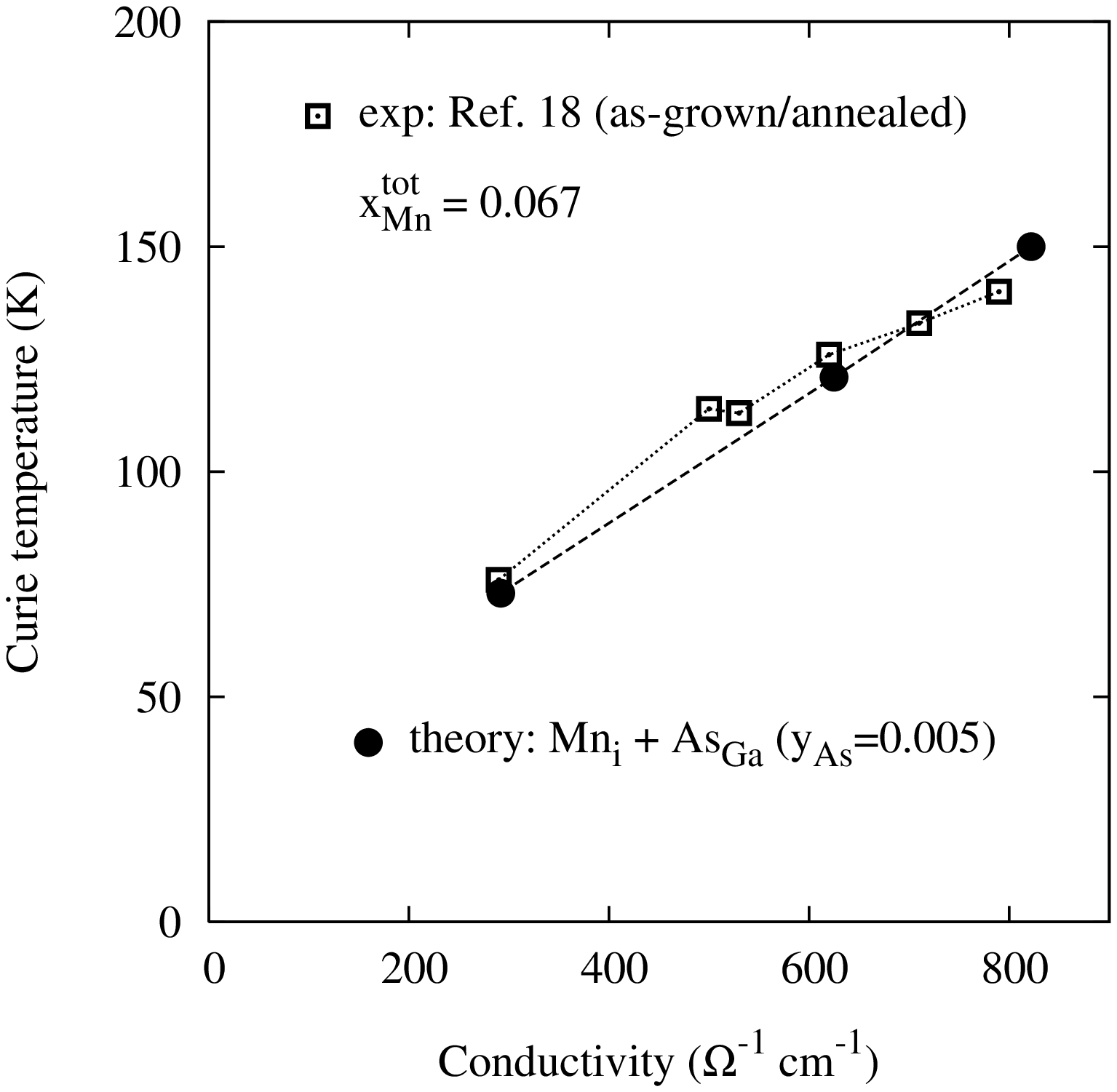}
\caption {T$_{c}$ vs conductivity relation assuming that
dominating compensating defects are Mn-interstitials but
there is present also a certain concentration $y_{\rm As}=0.005$ 
of As$_{\rm Ga}$-antisites. 
Experimental results \cite{tcexp} for as-grown and
annealed samples are also shown.
}
\label{f3}
\end{figure}


\newpage

\begin{thebibliography}{99}

\bibitem{review} T. Jungwirth, J. Sinova, J. Ma\v{s}ek, 
J. Ku\v{c}era, and A.H. MacDonald, 
Rev. Mod. Phys. {\bf 78}, 809 (2006).
\bibitem{edmonds} K.W. Edmonds, K.Y. Wang, R.P. Campion, 
A.C. Neumann, N.R.S. Farley, B.L. Gallagher, and C.T. Foxton,
Appl. Phys. Lett. {\bf 81}, 4991 (2002). 
\bibitem{dietl} T. Dietl, H. Ohno, and F. Matsukara, 
Phys. Rev. B {\bf 63}, 195205 (2001).
\bibitem{jungw} T. Jungwirth, M. Abolfath, J. Sinova, 
J. Ku\v{c}era, and A.H.MacDonald,
Appl. Phys. Lett. {\bf 81}, 4029 (2002).
\bibitem{lars} L. Bergqvist, O. Eriksson, J. Kudrnovsk\'y,
V. Drchal, P. Korzhavyi, and I. Turek,
Phys. Rev. Lett. {\bf 93}, 137202 (2004).
\bibitem{sato} K. Sato, W. Schwejka, P.H. Dederichs, and 
H. Katayama-Yoshida,
Phys. Rev. B {\bf 70}, 201202(R) (2004);
\bibitem{epl} G. Bouzerar, T. Ziman, and J. Kudrnovsk\'y,
Europhysics Lett. {\bf 69}, 812 (2005).
\bibitem{lie} A.I. Liechtenstein, M.I. Katsnelson, 
V.P. Antropov, and V.A. Gubanov, 
J. Magn. Magn. Mater. {\bf 67}, 65 (1987).
\bibitem{eirev} I. Turek, J. Kudrnovsk\'y, V. Drchal,  and P. Bruno,
Phil. Mag. {\bf 86}, 1713 (2006).
\bibitem{ourJ} J. Kudrnovsk\'y, I. Turek, V. Drchal, F. M\'aca,
P. Weinberger, and P. Bruno, Phys. Rev. B {\bf 69} 115208 (2004).
\bibitem{book} I. Turek, V. Drchal, J. Kudrnovsk\'y, 
M. \v{S}ob, and P. Weinberger, {\it Electronic Structure of 
Disordered Alloys, Surfaces and Interfaces} 
(Kluwer, Boston, 1997).
\bibitem{alm} G. Bouzerar, J. Kudrnovsk\'y, L. Bergqvist, and
P. Bruno, Phys. Rev. B {\bf 68}, 081203 (2003). 
\bibitem{mnias} G. Bouzerar, T. Ziman, and J. Kudrnovsk\'y,
Phys. Rev. B {\bf 72}, 125207 (2005).
\bibitem{mnint} J. Ma\v{s}ek and F. M\'aca, 
Phys. Rev. B {\bf 69}, 165212 (2004); 
R. Wu, Phys. Rev. Lett. {\bf 94}, 207201 (2005). 
\bibitem{trspd} M.E. Fisher and J.S. Langer,
Phys. Rev. Lett. {\bf 20}, 665 (1968); T. Kasuya and A. Kondo,
Solid State Commun. {\bf 14}, 253 (1974).
\bibitem{kglmto} I. Turek, J. Kudrnovsk\'y, V. Drchal, L. Szunyogh,
and P. Weinberger, Phys. Rev. B {\bf 65}, 125101 (2002).
\bibitem{trit} I. Turek, J. Kudrnovsk\'y, V. Drchal, and P. Weinberger,
J. Phys.: Condens. Matter {\bf 16}, S5607 (2004).
\bibitem{tcexp} Unpublished experimental data obtained for various
stages of annealing, from as-grown sample to almost perfectly annealed
sample, for (Ga,Mn)As alloy with nominal concentration of magnetic
atoms $x_{\rm Mn}=0.067$. The conductivity was measured at T=4.2~K. 
Further details can be found in Ref.~\onlinecite{edmonds}.

\end{thebibliography}
\end{document}